# Nucleosynthesis in the Cosmos:
# What we think we know and forthcoming questions

# Nucleosíntesis en el Cosmos:
# Lo que creemos que sabemos y las preguntas por venir

Salvador Galindo Uribarri[1] and Jorge L. Cervantes-Cota[1]

[1]Instituto Nacional de Investigaciones Nucleares.
Km. 36.5 Carretera México-Toluca, La Marquesa, Ocoyoacac, Estado de México. C.P. 52750, Mexico.

**Abstract:**

We present what we know on nucleosynthesis in the Universe and hypotheses that have been made in this regard. A brief description of the Universe's evolution during its different stages is offered, indicating which are the periods and mechanisms of element formation. A critical prospective on future research is formulated to validate, modify, or reject the hypotheses formulated. These will involve joint observations that encompass finer measurements of cosmic background radiation, galaxy clusters, and gravitational waves produced by neutron star collisions. The information thus obtained will be combined with restrictions given by theoretical models. Perhaps many current doubts will be clarified, but new questions will arise.

**Resumen:**

Se da una breve descripción de lo que creemos saber hasta el día de hoy sobre la nucleosíntesis de elementos en el universo. Por medio de una reseña cronológica se indican los periodos y mecanismos de su formación. Se formula una prospectiva crítica sobre investigaciones futuras que permitirán validar, modificar o rechazar hipótesis hechas, las cuales involucrarán observaciones conjuntas que abarcan medi- ciones más finas de radiación cósmica de fondo, cúmulos de galaxias y de ondas gravitacionales producidas por choques de estrellas de neutrones. La información obtenida se combinará con restricciones dadas por modelos teóricos. Tal vez, muchas dudas actuales serán aclaradas, pero nuevas preguntas surgirán.

**Palabras clave**: nucleosíntesis, elementos, orígenes cosmológicos

# Introduction

The atomic composition of the Universe is fairly simple. About 75% (by mass) of atomic matter is in the form of hydrogen, and almost all the rest is in the form of helium. All other elements contribute less than 2% of the universe baryonic mass. Why is it so? To answer this question, it is inevitable to go back to the origin of the Universe, since atoms could not exist under conditions that prevailed during the early moments of what is now known as the Big Bang. Bound to happen, nucleosynthesis of elements is tied to the evolution of our Universe.

This paper deals with how elements were created by physical processes during the course of cosmic evolution and on the doubts, assumptions and new questions that have arisen about the involved physical processes.

Our challenge is to write a short but inclusive narrative of these issues. For this purpose, we have divided our work into two parts. The first will deal with what we think we know about cosmic evolution and nucleosynthesis. In the second part we will make a brief prospective examination on those issues whose current validity has been questioned or require rethinking. Finally, we shall give an account of the new questions that have emerged.

## 1. What we think we know:

### 1.1 On the origin of the Universe

We know that the Universe is vast and ancient, with an age of around 13.8 billion years and a diameter of 93 billion light-years, equivalent to 880,000 quintillion kilometers (e.e., Gribbin, 2016). Indeed a mind-blowing vastness, and with us as observers in its center perceiving the horizon of astronomical events. However, we are not so unique, each point and every point of the Universe is its center, from which the same panorama is seen, from anywhere and in any direction, that is, the Universe is homogeneous and isotropic on cosmic scales, i.e. scales of roughly 150Mpc or more (e.g., Kasai, 1993).

This is the reason why what we uncover in the surroundings of our Solar System, which is immersed in a typical large galaxy, is the same as what is found in other typical galaxies in the Universe. Our Milky Way has about 100,000 million stars similar to our Sun. The Earth orbits around a typical star, within a typical galaxy of the Universe. For this reason we know that elements of the periodic table swarm everywhere in the Universe, but they are distributed under a logic that is regulated by the origin and evolution of the Universe (e.g., Hogan 2000).

## 1.2 On the early evolution of the Universe

Today it is widely accepted that The Universe was born in the so-called Big Bang. It was born from a very small initial region, with extreme conditions of high temperature, density and pressure. From this tiny region the expansion of the Universe arises. We do not know exactly the details of this beginning, but taking into account the laws of physics, and extrapolating them to their limit of validity, with Planck conditions, they lead us to deduce that the Universe had a size of $10^{-33}$ cm (one thousandth of a nonillionth of a centimeter!), at a starting time of $10^{-43}$ s (one tenth of tredecillionth of a second), with a density of $10^{92}$ gr/cm$^3$. These conditions have only been experienced once in time, at the very beginning of the Big Bang.

Without doubt, with these physical conditions standard matter cannot exist, for under these conditions standard matter was broken up into fields, whose characterization we have not been able to decipher in detail to date. But we know that those fields can be described, within the context of quantum field theory, in its semi-classical approach (e.g., Rajaraman, 1975).

Here, fields existed in a geometry that is described, from then until now, with Einstein's theory of General Relativity (e.g., Thiemann, 2008). This theory relates geometry to matter, and under these conditions fields cause space to expand at unusual rates, in such a way that the first stage of the Universe suffers an accelerated expansion known as "Inflation", growing $10^{28}$ times in an epoch that lasted from $10^{-36}$ seconds, after the conjectured Big Bang singularity, to sometime between $10^{-33}$ and $10^{-32}$ seconds after the singularity (Tsukijawa, 2003). At the end of the inflationary stage, the Universe had a size of one hundred thousandth of a centimeter, a size already very large considering its initial extent. Eventually, after new phase transitions of the fields (e.g. Higgs) and growth of the Universe, its material content is described by the standard model of particle physics, that is a unified scheme that pretends to accommodate the known particles (bosons and fermions) and interactions in a unified framework. In their evolution, fields shall form quarks, electrons, neutrinos, and more photons, that later, in subsequent times will form protons, neutrons; basis for the formation of the chemical elements of the Universe. We will describe this at its time.

From its very beginning the Universe is thought to be expanding, and it continues until now, but of course with different expansion rates. This pace is determined by the dominant energy content in cosmos. In the course of the early Universe, different events happen, in which a series of field transformations occurred. The detailed description of the cosmic, thermal evolution for different particle types, depending on their masses, cross-sections, etc., is well described in many textbooks. We notice here that, as the Universe cools down a series of spontaneous symmetry–breaking phase transitions were expected to occur. The type and/or nature of these transitions depend on the specific particle physics theory considered. Among the most popular ones are Grand Unification Theories (GUT's), which bring together, in a single theoretical scheme, all known interactions except of gravity. One could also be more modest and just consider the standard model of particle physics or some extensions of it. Ultimately, one should settle, in constructing a cosmological theory, up to which energy scale one wants to describe physics. For instance, at a temperature between $10^{14}$ GeV to $10^{16}$ GeV (where we use units in which the Boltzmann constant and speed of light are unity) the transition of the SU(5) GUT should take place, if this theory is valid, in which a type of Higgs field breaks this symmetry to SU(3)$_C$ ×SU(2)$_W$ ×U(1)$_{HC}$, a process through which some bosons acquire their masses. Due to the gauge symmetry, there are: color (C), weak (W) and hypercharge (HC) conservation, as the sub-index indicates.

Later on, when the Universe evolved to around 250 GeV the electroweak phase transition took place, in which the known Higgs field breaks the symmetry SU(3)$_C$ ×SU(2)$_W$ ×U(1)$_{HC}$ to SU(3)$_C$ ×U(1)$_{EM}$; through this second breaking also the fermions acquire their masses. At this stage, there are only color and electromagnetic charge conservation, due to the gauge symmetry. During the Quark epoch, the Universe was filled with hot quark–gluon plasma, a "quark soup". From this point onwards the physics of the early Universe is much better understood, and the energies involved in the Quark epoch are directly accessible in particle physics experiments such as the Large Hadron Collider.

The Quark epoch began roughly $10^{-12}$ seconds after the Big Bang. This was the period in the evolution of the early Universe when the fundamental interactions of gravitation, electromagnetism, the strong interaction and the weak interaction had taken their present forms, but the temperature of the Universe was still too high to allow quarks to bind together to form hadrons (e.g., Petter, 2013). Afterwards, at a temperature of a few hundred of MeV the Universe should undergo a transition associated to the chiral symmetry–breaking and color confinement, from which baryons and mesons are formed out of quarks. The so called "Hadron epoch", where the Universe keeps on cooling and quarks are bound into: hadrons, protons, and neutrons. A slight matter-antimatter asymmetry possibly originated in earlier times results in a removal of anti-hadrons (e.g., Sather, 1996). If this would not be the case, matter and anti-matter annihilate, and the universe composition would only be photons. Our own existence prevails that!

Subsequently, at approximately 10 MeV, at a thousand of a second, the synthesis of light elements (nucleosynthesis) begins, when most of the today observed hydrogen, helium, and some other light elements abundances were produced. The nucleosynthesis represents the earliest scenario tested in the standard model of cosmology. Later, when the Universe was one second old, Neutrinos cease interacting with leptonic matter (electrons, positrons) and leptons and antileptons remain in thermal equilibrium. Recall that to maintain thermal equilibrium the rate of interaction of particles (its cross section) should happen faster than the expansion of the Universe. These decoupled neutrinos should form the cosmic neutrino background (CνB), yet undetected. If primordial black holes exist, they could also be formed at about one second of cosmic time. The Universe is 10 light-years across and is ten seconds old. Composite subatomic particles emerge—including protons and neutrons—and from about 2 minutes onwards, conditions are suitable for nucleosynthesis.

## 1.3 On Big Bang nucleosynthesis

The Universe continues to expand, and as it expands it further cools down with its contents of protons, neutrons, photons, and decoupled neutrinos. And although there are more entities present, at that moment, they do not play a relevant role in the creation of the lightest elements nuclei.

The most accepted theory to explain the creation light elements nuclei, in this early stage of the Universe, is the so-called Standard Big Bang Nucleosynthesis (SBBN) (e.g., Sarkar, 1996). This theory is well understood in the context of the standard model of particle physics (Oerter, 2006). The SBBN model contains few parameters such as the baryon-to-photon ratio $\eta = n_b/n_\gamma$, the neutron decay time $\tau_n$, and the number of neutrino families $N_\nu$ (e.g., Bertulani. 2016). These parameters are inferred from direct observations.

The parameter $\eta$ links the baryon density of the Universe $\Omega_0$ to the Hubble dimensionless parameter $h$ defined as $H_0 = 100\ h$ km/s/Mpc (the index '0' denotes present time), through the relation $\Omega_0 h^2 \cong \eta/(273 \times 10^{-10})$. The value of $\eta$ can be independently inferred from the observed anisotropies of the cosmic microwave radiation (CMB) (P.A.R. 2016). That is the time when photons decoupled and began streaming freely in the Universe. As for the number $N_\nu$ of neutrino families, the Large Electron Positron collider (LEP) experiments at CERN deduce it to be three (ALEPH, 2006); measurements of the CMB anisotropies are also in consistency with that number of neutrino families. Neutron lifetime measurements have obtained a value of $\tau_n \cong 880.2 \pm 1.0$ s (Patrignani, 2016).

The SBBN model predictions also depend on the nuclear reaction network (figure 1) and values of nuclear cross sections. The main reactions involved in the Big Bang Nucleosynthesis period are shown in table 1. Reactions occur among neutrons (n), protons (p), deuterium (d), tritium (t), and the first created elements: H, He, Li and Be

| Table 1. Reactions of relevance for SBBN from the NACRE-II (Xu, 2013) | | | |
|---|---|---|---|
| 1) n↔p | 2) p(n,γ)d | 3) d(p,γ)$^3$He | 4) d(d,n)$^3$He |
| 5) d(d,p)t | 6) t(d,n)$^4$He | 7) t(α,γ)$^7$Li | 8) $^3$He (n,p)t |
| 9) $^3$He(d,p)$^4$He | 10) $^3$He(α,γ)$^7$Be | 11) $^7$Li(p,α)$^4$He | 12) $^7$Be(n,p)$^7$Li |

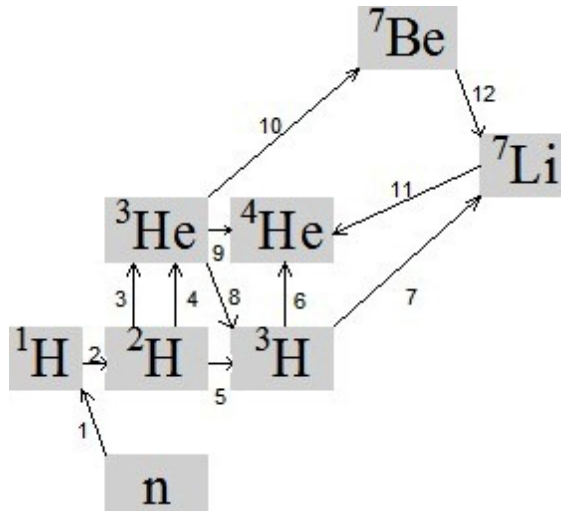

**Figure 1. Minimal network of SBBN:** see Table 1 for the labeled reactions (figure credit, own source)

At the end SBBN predicts that the Universe is composed of about 75% of Hydrogen and 25% of $^4$He and small amounts of D, $^3$He, $^7$Li and $^6$Li. In addition, insignificant amounts of $^3$H, $^7$Be and $^8$Be were formed, but these are unstable and are quickly lost again (e.g., Karki, 2010). Figure 2 shows the periodic table of elements about 250 seconds after the Big Bang.

**Figure 2. The 18-column form of the periodic table of elements 250 seconds after the Big Bang:** Numbers at top of each the column indicate the group. Numbers inside boxes indicate atomic number. The first primordial 3 elements (H, He, and Li) are enclosed in their proper boxes. Insignificant amounts of $^7$Be and $^8$Be were formed but due to their short lifetime they decayed (figure credit, own source).

The light element abundances as predicted by the SBBN model ("Schramm Plot") are shown in figure 3 compared to values obtained from WMAP (Wilkinson Microwave Anisotropy Probe) observations of the Cosmic Microwave Background (CMB).

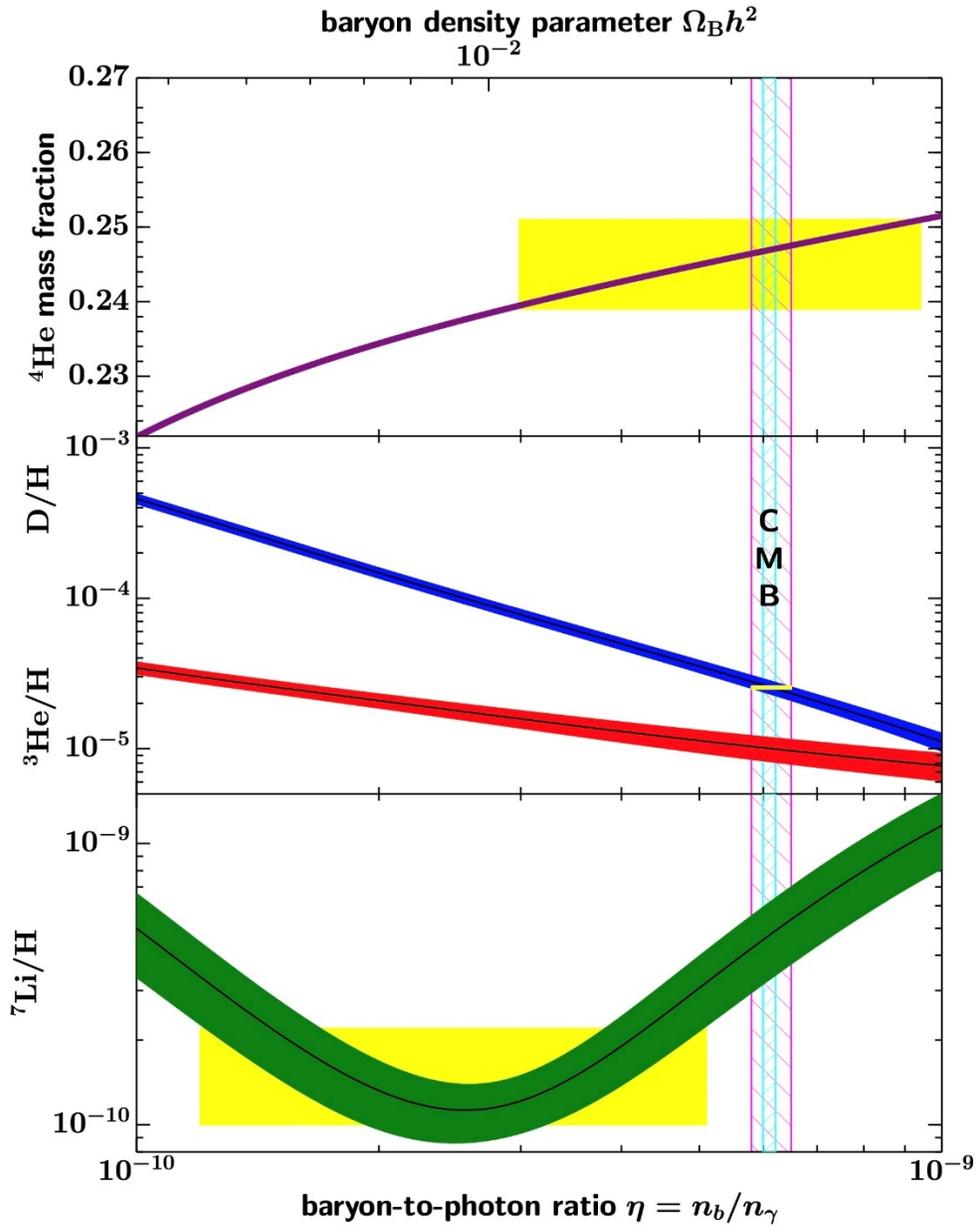

**Figure 3. "Schramm plot":** This figure depicts primordial abundances of $^4$He, D, $^3$He, and $^7$Li as a function of cosmic baryon content from SBBN predictions. CMB predictions of $^7$Li (narrow vertical bands, at 95% confidence) and the SBBN D + $^4$He concordance range (wider vertical bands, at 95% CL) should overlap with the observed light element abundances (boxes) to be in agreement. This occurs in $^4$He and is well constrained in D, but is not the case for $^7$Li, where the observed Li observations lie a factor of 3 to 4 below the SBBN+WMAP prediction (credit NASA/WMAP Science Team).

## 1.4 On Recombination and photon decoupling

After an initial period of big bang nucleosynthesis in the first twenty minutes of the life of the cosmos, the Universe expanded, cooled and the production of baryonic matter ceased. Until then, the expanding Universe had been a plasma of photons and nuclei; an optically opaque plasma where photons were not able to freely travel through the Universe, as they were constantly scattered off. This Thomson scattering process caused a loss of information, and there is therefore a "photon barrier" that prevents us from using photons directly to learn about the Universe at larger redshifts (e.g., Longair, 2008). But when the temperature fell to about 3000 °K, charged electrons and $^1H$, $^2H$, $^3He$, $^4He$ and $^7Li$ nuclei first became bound to form electrically neutral atoms. This period is called recombination epoch and occurred after about 380,000 years after the Big Bang (Tanabashi, 2018). Shortly after recombination, the photon mean free path became larger than the Hubble length, and photons traveled freely without interacting with matter (e.g., Padmanabham, 1993). The latter process is called "photon decoupling" or last scattering surface. Then the Universe became optically clear, transparent. In fact, the light from that epoch is now detected as CMB at T=2.725 °K.

## 1.5 On Dark Ages, Reionization and advent of Stelliferous era

After recombination and decoupling, the Universe had cooled enough to allow light to travel long distances, but there were no light-producing structures such as stars and galaxies. This period, known as the "dark Ages", began around 380,000 years after the Big Bang. During the dark Ages, the temperature of the Universe cooled from some 3000 °K to about 60°K. The beginning of the end of dark Ages commenced when the first stars formed, then light was emitted by a new source. The first generation of stars, known as Population III stars (or Pop III) [1], began to "light up" within a few hundred million years after the Big Bang (e. g., Yoshida 2006), which in cosmological terms is not a terribly long time.

Pop III stars not only produced visible light, but also copious amounts of UV photons to reionize the universe (e.g., Venkatesan , 2003, Venkatesan, 2003a). These stars were the first light sources in the Universe after recombination. To the same extent as these stars emerged, dark Ages gradually ended, marking the advent of the "stelliferous" era. Since this activity was gradual, the dark Ages only lasted entirely from an age of 380,000 years to around 500 million years after the Big Bang.

Here is appropriate and timely to mention that Pop III stars launched the critical evolution of the Universe from a homogeneous, simple entity to a highly structured, complex one at the end of the cosmic dark ages. Universe structures may have begun to appear from around 150 million years from the Big Bang, and early galaxies emerged from around 380 to 700 million years, and our Universe began taking its present appearance.

---

[1] For historical reasons the first stars in the universe were called Pop III stars because previously two other younger star groups were branded Pop I and Pop II.

Until recently, we didn´t have separate observations of very early individual stars. But in 2015, an observation of a Pop III star was confirmed (Sobral, 2015). Nowadays Pop III computational models are currently tested and our understanding of their formation and evolution is expected to improve with upcoming more powerful ground and space telescopes.

## 1.6 On Pop III stars

Population III stars are responsible for turning the few light elements that were formed in the Big Bang (hydrogen, helium and small amounts of lithium) into many heavier elements[2,3]. The central question about Pop III stars is how massive they typically were. Results from numerical simulations of the collapse and fragmentation of primordial gas clouds imply that these stars were mostly very massive, with typical masses M ≥ 100 $M_{sun}$ (Bromm, 2002, Abel, 2002).

This huge mass value is consistent with the characteristic fragmentation scale, given approximately by the Jeans mass Mj which is proportional to the square of the gas temperature and inversely proportional to the square root of the gas pressure. The first star-forming systems would have had pressures similar to those of present-day molecular clouds. But because the temperatures of the first collapsing gas clumps were almost 30 times higher than those of molecular clouds, their Jeans mass would have been almost 1,000 times larger i.e., Mj ~ $10^3$ $M_{sun}$ (Clarke, 2007). The higher temperature of the $H_2$ clump is explained by the microphysics of molecular hydrogen ($H_2$) cooling, which is only possible through rotational and vibrational transitions of $H_2$ molecules.

However, having constrained the characteristic mass scale, still leaves undetermined the overall range of Pop III stellar masses (e.g., Omukai, 2003). In addition, it is presently not known whether binaries or, more generally, clusters for these zero-metallicity stars, can form.

## 1.7 On Pop III nucleosynthesis

Pop III were the earliest and only stars to appear out of the primordial mixture of H/He, still a zero metallicity gas. Their arrival launched the critical evolution from a homogeneous, simple universe to a highly structured, complex one at the end of the cosmic dark Ages. This means that early Universe was significantly less complex than the present one. Stars evolution was triggered by gravitation in the process of collapse and star formation. From a theoretical perspective this circumstance presents advantages for carrying out high-level numerical simulations using our present knowledge on nuclear reactions and fluid mechanics (e.g., Klapp, 2007).

---

[2] Astrophysics nomenclature considers any element heavier than helium to be a "metal", including chemical non-metals such as oxygen.
[3] Pop III stars had very low metal content, Pop II are older stars of low metallicity and recent stars (i.e. Pop I stars) are of high metallicity.

We have just mention that the very first Pop III stars were initially composed of nearly pure hydrogen and a fraction of helium. At a certain time, gravitational collapse in denser regions of a primordial hydrogen and helium cloud caused their heating until the cloud became so hot in its center that nuclear burning of H into He began. Consequently proton-proton (pp) chains must be invoked to synthesize $^4$He in a gas mostly composed of hydrogen and to some minimal extent helium. The pp-chains play a key role to supply the helium needed to produce some of the isotopes involved in the subsequent CNO cycles. Once the ignition started in these H-burning stars, the pp-chains and the CNO cycles ran simultaneously. The net result of the CNO cycles is to produce $^4$He from $^1$H. Additionally, the conversion of C, N and O isotopes transforms them mostly into $^{14}$N, since the $^{14}$N(p, γ)$^{15}$O reaction is slower than other involved reactions. Figure 4 shows the full CNO tetra-cycle.

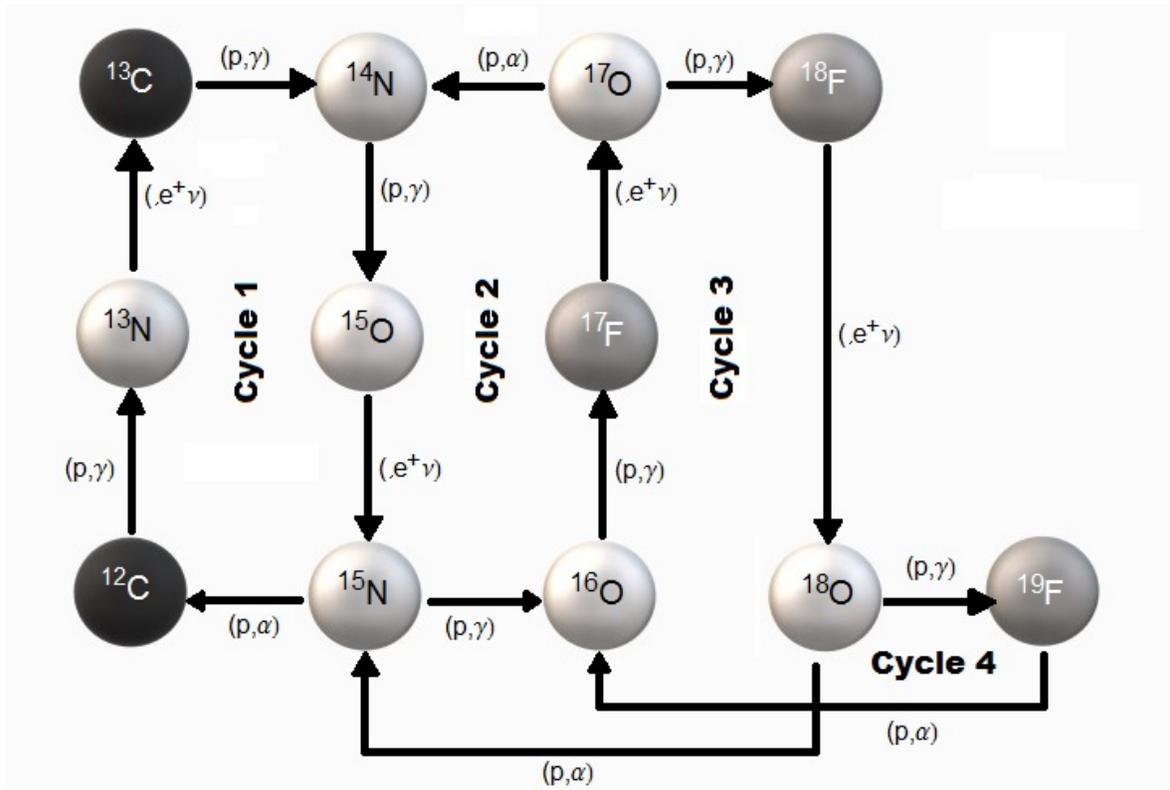

**Figure 4. Full CNO tetra-cycle:** Using the C, N, and O isotopes as catalysts during H burning: produces $^{14}$N from initial $^{12}$C and $^{16}$O (figure credit, own source).

After central H is depleted, the star contracts until higher temperatures are reached. Upon reaching a temperature that allows overcoming the higher Coulomb barriers, the burning of He gets underway. He-burning reactions gradually convert $^4$He into $^{12}$C, $^{16}$O, and so forth. The core of fusion reactions by which 3$^4$He → $^{12}$C + γ is the transitory creation of $^8$Be from two alpha particles. This reaction is called "triple alpha" (3α) and is strongly temperature dependent.

When sufficient amounts of $^{12}$C have been built up by the 3α reaction, further α captures can occur and the nuclei $^{16}$O, $^{20}$Ne... are successively produced. The $^{12}$C(α, γ)$^{16}$O reaction is non-resonant and occurs simultaneously with the 3α reactions. In fact, during He-burning, the whole set of reactions occur simultaneously[4]. As temperatures raise up to 1.5 x 10$^8$ K, the main He-burning reactions are: 2 $^4$He(α, γ)$^{12}$C and $^{12}$C(α, γ)$^{16}$O. The $^{16}$O(α, γ)$^{20}$Ne reaction also takes place during He-burning but at a much lower rate than $^{12}$C(α, γ)$^{16}$O.

The star evolution proceeds by the process of successive burning of the elements to produce nuclei with higher and higher binding energies. For Pop III massive stars, the sequence of burning runs through He-burning to form carbon, carbon and oxygen burning to produce silicon which can eventually be burned through to iron peak elements. These processes can be summarized as:

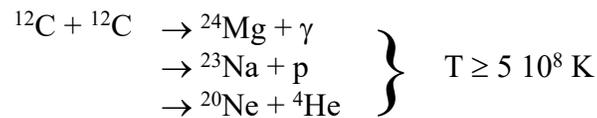

$$\left. \begin{array}{ll} ^{12}\text{C} + ^{12}\text{C} & \rightarrow ^{24}\text{Mg} + \gamma \\ & \rightarrow ^{23}\text{Na} + p \\ & \rightarrow ^{20}\text{Ne} + ^4\text{He} \end{array} \right\} \quad T \geq 5\ 10^8\ \text{K}$$

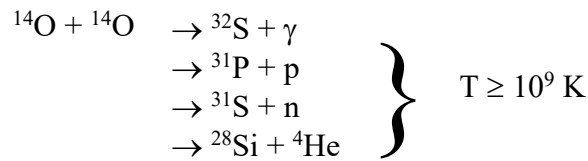

$$\left. \begin{array}{ll} ^{14}\text{O} + ^{14}\text{O} & \rightarrow ^{32}\text{S} + \gamma \\ & \rightarrow ^{31}\text{P} + p \\ & \rightarrow ^{31}\text{S} + n \\ & \rightarrow ^{28}\text{Si} + ^4\text{He} \end{array} \right\} \quad T \geq 10^9\ \text{K}$$

In case of Silicon burning, which begins at a temperature of 2 10$^9$ K, the reactions proceed slightly differently because the high energy gamma rays remove protons and $^4$He particles from the silicon nuclei and the heavier elements are thus synthetized by the addition of $^4$He nuclei through reactions which can be schematically written as

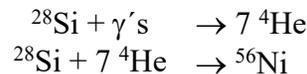

$$^{28}\text{Si} + \gamma\text{'s} \rightarrow 7\ ^4\text{He}$$
$$^{28}\text{Si} + 7\ ^4\text{He} \rightarrow ^{56}\text{Ni}$$

It is therefore predictable that in the final stages of the evolution of these massive stars, the star will adopt an "onion skin" structure with a central core of iron peak elements and successive surrounding shells of silicone, carbon oxygen, helium and hydrogen (see figure 5)

---

[4] Note: Reaction A + p⟶ B + x can also be written as A(p,x)B; the latter notation is generally adopted in this text, but the former only occasionally.

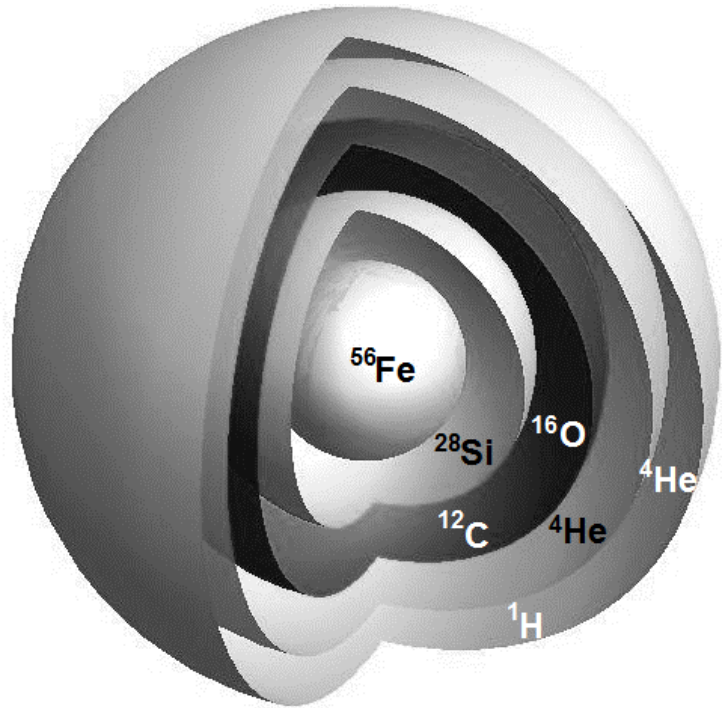

**Figure 5. Hypothetical inner structure of a massive star:** At a late evolutionary stage, the star consists of layers with different composition separated by nuclear burning shells (figure credit, own source)

Some of the nucleosynthesis products of Pop III stars may have been dispersed into the interstellar gas during their lifetime through mass loss episodes such as a pair instability supernova (PISN) for stellar masses between $140 - 260$ $M_{sun}$. Upon consuming its fuel after a lifetime of only about $3 \times 10^6$ yr, a Pop III star in the mass range from $10-140$ $M_{sun}$, will undergo Fe core collapse at the end of their evolution becoming Type II or Ib/c supernovae (Fryer, 2001). In any of the mentioned cases the explosions enrich the intergalactic medium with heavy elements. For masses above $\sim 260$ $M_{sun}$, a Pop III star is predicted to collapse entirely into a massive black hole without any concomitant metal ejection (Ohkubo et al. 2009). That said, many details regarding, for example, the mass ranges and explosion energies involved in the evolution of Pop III are still being debated.

The supernovae expelled the elements produced during nucleosynthesis up to the iron peak together with: neutrons, helium, and elements which were formed mostly during the late stages in the evolution of the star in an environment with high neutron densities, in excess of $10^6$ free-neutrons per cubic centimeter. In this process, a free neutron might be captured onto a seed nucleus. The subsequent products depend upon whether or not the nucleus formed has time to decay before the addition of further neutrons. At the end it results in a heavier, radioactive nucleus that subsequently decays into a stable heavy species. This is the so-called slow neutron-capture process, or s-process.

But the s-process accounts for the formation of only about half of the isotopes beyond iron. Creating the other half requires a rapid capture sequence, the r-process, and a density of greater than $10^{20}$ neutrons/cm$^3$ that can bombard seed nuclei. We shall go into the r- process later in this paper.

Figure 6 shows the periodic table of elements 100 million years after the Big Bang.

**Figure 6. The 18-column form of the periodic table of elements 100 million years after the Big Bang:** Slight amounts of $^3$He, together with lithium, beryllium, and boron are formed by cosmic ray spallation. Elements beyond the iron peak are due to slow neutron captures s-process (figure credit, own source).

## 1.8 On the end of Pop III era and the start of Pop II and Pop I

The epoch of Population III star formation eventually terminated. Pop III type stars ceased to be created due to two different effects, the first being radiative and the second chemical in nature (e.g., Mackey, 2003). The radiative effect that inhibited Pop III stars formation involves the soft UV photons (UVB) produced by the first stars. This radiation photo-dissociated the rather fragile H$_2$ molecules in the surrounding gas, thus suppressing their only corresponding cooling agent, this is molecular H$_2$. (e.g., Haiman, 2000) Massive Population III stars could then no longer form.

The second inhibiting effect that was important in terminating the epoch of Pop III star formation is chemical in nature, or, more precisely: it is due to the enrichment of the primordial gas with the heavy elements dispersed by the first SNe. Numerical simulations of the fragmentation process have shown that lower mass stars can only form out of gas that was already pre-enriched to a level in excess of the 'critical metallicity', estimated to be of order $10^{-4}$ to $10^{-3}$ the solar value (Bromm, 2001). Depending on the nature of the first SNe explosions, and in particular on how efficiently and widespread the mixing of the metal-enriched ejecta proceeds, the cosmic star formation at some point underwent a fundamental transition from an early high-mass (Population III) dominated mode to one dominated by lower mass stars (Population II).

The first Pop II stars are thought to be formed from a low-energy (~1051 erg) type II supernova of a Pop III ranging between 40 and 60 Msun (Keller, 2014) and not PISN events.

The reason for this conviction is that violent PISNs are in fact likely to disrupt the hosting halo inhibiting the formation of a second generation of stars (Seifried, 2014). Elliptical galaxies are now observed to consist primarily of old, red, Pop II stars.

## 1.9 On Reionization and formation of structures.

Reionization is the transition from the neutral universe to the completely ionized universe that is: the present one. This transition started off as a result of the formation of Pop III stars and the ionizing radiation they emitted. As we have just mentioned, the production of UV radiation in Pop III stars inhibited their subsequent own regenerations, but we must also point out that the mentioned radiation gradually ionized the Universe, perhaps patchy way, from redshifts $z=20$ to $z=10$, and after this time the universe became fully re-ionized (Gnedin, 2004).

At first, Pop III stars would simply form pockets of ionized gas around them which would then shrink back as the atoms recombined. Eventually, however, with the continued formation of stars and radiation emitted from active galactic nuclei, the universe has become fully ionized.

The very first metals that were expelled from the supernovae of these Pop III stars and the remnant black holes they leaved behind may have been the seeds for the supermassive black holes found in the centers of galaxy clusters today.

In the aftermath of violent events that suppressed the renewal of Pop III stars and favored the emergence of Pop II stars, the ejected SNe debris and dust grains from both -Pop III and II- gradually began to cool the interstellar gas down to the CMB temperature, favoring the nucleation of the first high metallicity Pop I stars. The metals ejected from previous generations of stars and explosive events possibly induced fragmentation during high-redshift structure formation, and therefore, determined the mass scale of new generations of stars (Bovino, 2014). In addition, it is considered that the mentioned fragmentation in turn stimulated the emergence of today´s stars close arrangements.

Under the current paradigm of hierarchical structure formation as dictated by Lambda-Cold Dark Matter (ΛCDM) cosmology, smaller objects build up to form larger ones. The modern ΛCDM model is successful at predicting the observed structure of our Universe, that is: large-scale distribution of galaxies, clusters and voids; but on the scale of individual galaxies there are many complications due to highly nonlinear processes involving baryonic physics, gas heating and cooling, star formation and feedback.

For example, there is a mismatch between observed dwarf galaxy numbers and numerical simulations that predict the evolution of matter distribution in the universe. This is the so called "the Dwarf galaxy problem" also known as the missing satellites problem. That is, the number of small ultrafaint dwarf (UFD) galaxies is orders of magnitude lower than expected from simulation (Simon, 2019). Figure 7 shows the Milky Way and UFD satellites.

There are two proposals to solve the "Dwarf galaxy problem". The first is that small-sized clumps of dark matter are unable to retain baryonic matter. Support for this proposal is the discovery in 2007 of 8 ultra-faint Milky Way dwarf satellites UFD of which 6 were 99.9% dark matter (Simon, 2007). The second proposal is that their proximity to a large galaxy tears them apart, tidally stripping them (Simon, 2017). But whatever the reason for the scarcity of UFD galaxies, their importance in the study of nucleosynthesis lies in the fact that they have served to discard supernova explosions as the main event in the production of r-process elements, as we shall see.

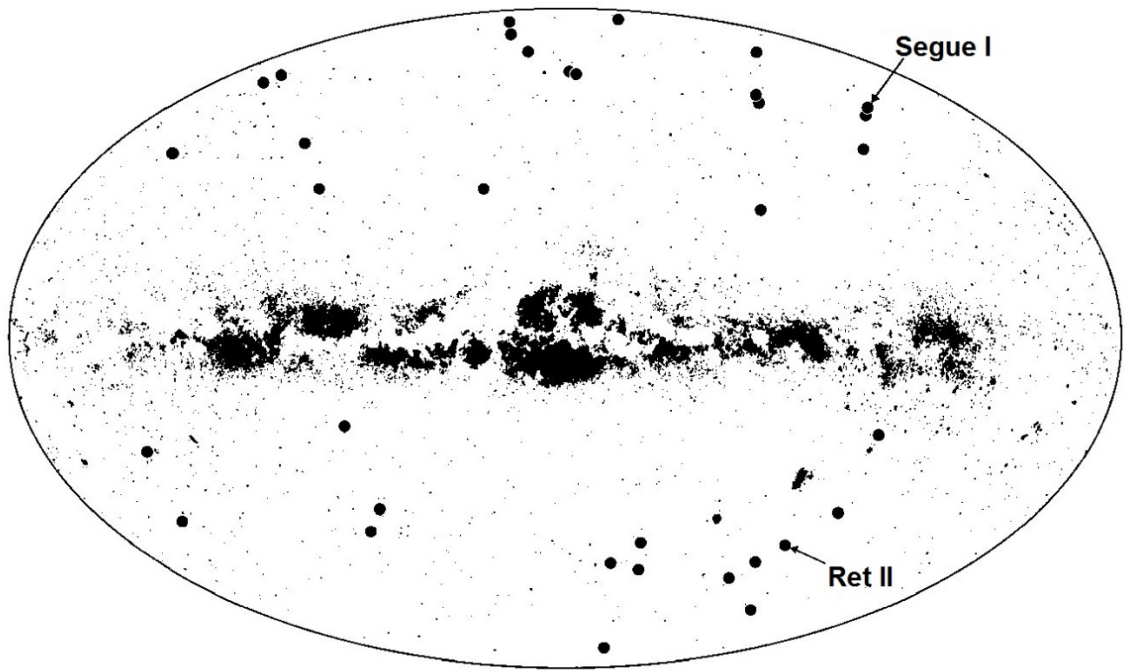

**Figure 7. Map of Milky Way UFD satellites (dots):** Segue 1 and Reticulum II (Ret II), both mentioned in the article, are specifically identified (credit, Helmut Jerjen, the European Space Observatory ESO).

## 1.10 On r-process Nucleosynthesis

We have already mentioned that s-process is responsible for the nucleosynthesis of only about half of the isotopes beyond iron and that the other half requires a rapid capture sequence of neutrons in a density environment of greater than $10^{20}$ neutrons/cm$^3$ i.e., the r-process. But the question that immediately arises is: where does the rapid neutron capture process occur? For a long time the supernova-explosion scenario had been favored as the principal event where the r-process took place. The reason was simply that during a SN explosion numerous neutrons are liberated. This neutron-rich environment was a potential candidate where nucleosynthesis of heavy elements occurred. However, several observational evidences ruled out this possibility, including observations made in the above mentioned UFD galaxies.

To explain the relevance of UFD galaxies to rule out the SNe scenario, we must point out some of their characteristics. The first is that these are galaxies that consist of very old stars implying a very early formation of these structures. The second is that most UFDs contain only a few thousand stars—much fewer than typical star clusters and the third, its stars have a high metallicity, particularly of elements produced by the r-process (e.g., Simon 2019).

In 2016 an UFD Milky Way satellite galaxy called Reticulum II (Ret II), provided evidence that the supernova-explosion scenario could not be the main mechanism for the production of the heaviest elements (Ji, 2016). In this UFD galaxy elements produced by r-process are found in excess in some of its stars. This enhancement would have required hundreds of thousands of supernovae explosions, but a small UFD galaxy simply does not have sufficient binding mass to have survived such a large number of cataclysms. Although it was expected some SNe had exploded, a small number would not produce significant amounts of r-process elements. So where do high neutron fluxes -necessary for the r-process heavy-element yield - derive? The next alternative candidates were then, neutron-star mergers

It turns out that once the first generation of stars in a system like Ret II or Segue 1 exploded injecting energy into the system, these small galaxies needed about 100 million years to cool sufficiently for another round of star formations. That's just enough the time for a pair a neutron stars in a binary system, to spiral in toward each other and merge. The composition of the stars in Ret II and other recently found UFD galaxies strongly suggested that neutron-star mergers are the universe's way to make elements such as gold and platinum.

This neutron star merger nucleosynthesis scenario was confirmed by striking observations reported in October 2017. This event marked the beginning of the multi-messenger astronomy. Most of what is known about the universe comes from observations of electromagnetic radiation. However, there are other "cosmic messengers." Gravity waves are disturbances in space-time that can be detected by very large laser interferometers such as LIGO (e.g, Cervantes-Cota 2017).

On 17 August 2017, LIGO/Virgo collaboration detected a pulse of gravitational waves, named GW170817, associated with the merger of two neutron stars in NGC 4993, an elliptical galaxy in the constellation Hydra. GW170817. This event also seemed related to a short ($\approx$2 second long) gamma-ray burst, GRB 170817A, first detected 1.7 seconds after the GW merger signal, and a visible light observational event first observed 11 hours afterwards, pointing to a kilonova event (SSS17a). The associated weeks-long outbursts of electromagnetic radiation were spectroscopically analyzed confirming the suspected nucleosynthesis of the heaviest elements (Arcavi, 2017).

On 25 April 2019 gravitational waves from a second merger of two neutron stars was observed by the LIGO Livingston detector (Abbott, 2020). This second event took place about 500 million light-years away, which also confirm that neutron mergers in the past were not unusual, spreading heavy nuclei to the interstellar medium and thus explaining the presence of heavy metals in old UFD galaxies. Figure 8 shows how the periodic table of elements looked 200 million years after the Big Bang.

**Figure 8. The 18-column form of the periodic table of elements 200 million years after the Big Bang:** Slight amounts of $^3$He, together with lithium, beryllium, and boron are formed by cosmic ray spallation. Elements beyond the iron peak are due to slow neutron captures s-process. Elements beyond are due to rapid neutron captures r-process (figure credit, own source).

Now it is up to us to see how the nucleosynthesis of the elements and their propagation in the interstellar medium, are concatenated with the evolution and explosive end of the stars. Next we will make a brief synthesis about it.

## 1.11 On stars fates and nucleosynthesis

As stars evolve, their final destinies are determined by their initial masses. We have already mentioned that those which are massive (at least 8 times the solar mass) end up in a supernova Type II explosion (e.g., Gilmore, 2004), ejecting debris of newly created elements (e.g. Bliss, 2018) and as a remnant either a black hole or a neutron star is left (see figure 9.a). The latter neutron star might merge into another neutron star (see figure 9.b), producing new chemical elements via an r-process (e.g., Pian, 2017).

For middle-size stars (less than 8 solar masses), when they have consumed all their fuels (hydrogen, helium, carbon and oxygen, in that order) their fate is to swell, and thus becoming red giant stars. Later they will blow off their envelopes, thus enriching the interstellar media with elements up to the iron peak, leaving behind a compact white dwarf remnant (Figure 9.c).

If such a remnant happens to be orbiting in a binary pair, then it is possible for the white dwarf star to accrete matter from its partner. This partner can be anything from a giant star to an even smaller white dwarf (figure 8.d). Once a critical amount of matter is acquired by the dwarf, the star reaches a high temperature that causes the ignition of explosive nuclear burning reactions (Mazzali, 2007). This critical amount of matter is known as the Chandrasekhar mass ($M_{Ch} \approx 1.4\ M_{sun}$). When the white dwarf reaches this mass it explodes leaving no remnant, producing a Type Ia supernova (SNIa) (Nomoto, 1984). At high stellar material densities, burning yields nuclear statistical equilibrium isotopes, in particular radioactive $^{56}Ni$ which decays to $^{56}Co$ and $^{56}Fe$ making the SN bright (Kuchner, 1994). At lower densities intermediate mass elements are synthesised. Both groups of elements are observed in the optical spectra of SNe type Ia (Branch, 1985).

If the mass acquired by the white dwarf is less than the Chandrasekhar mass it can still go through a transient astronomical event known as "Nova", that causes the sudden appearance of a bright, apparently "new" star that slowly fades over several weeks or months, see figure 9.d (e.g., Prialnik, 2001). When the small mass white dwarf is close enough to its companion partner, the drawing accreted matter might form a dense but shallow atmosphere onto the surface of the white dwarf (Yoon, 2004). This atmosphere, mostly consisting of hydrogen, is thermally heated by the hot white dwarf and eventually reaches a critical temperature causing ignition of rapid runaway fusion. Recurrent nova processes may be repetitive because the companion star can again feed the dense atmosphere of the white dwarf.

For very low mass stars whose final mass just before death is less than about 1.4 $M_{sun}$ (Chandrasekar limit), which is believed to be the case for most stars in the universe, their cores collapse (figure 9.e). Then their cores continue to shrink until they reach a very high density becoming white dwarfs achieving equilibrium forming an electron degenerate gas. They can still radiate light. After many billions of years, white dwarfs will no longer radiate light and in the very far future they will then turn to be black dwarfs—cold stellar corpses with the mass of a star and the size of a planet composed mostly of carbon, oxygen, and neon. However, the time required for a white dwarf to reach this state is calculated to be longer than the current age of the universe (13.8 billion years), so no black dwarfs are expected to exist in the universe now (Heger, 2003).

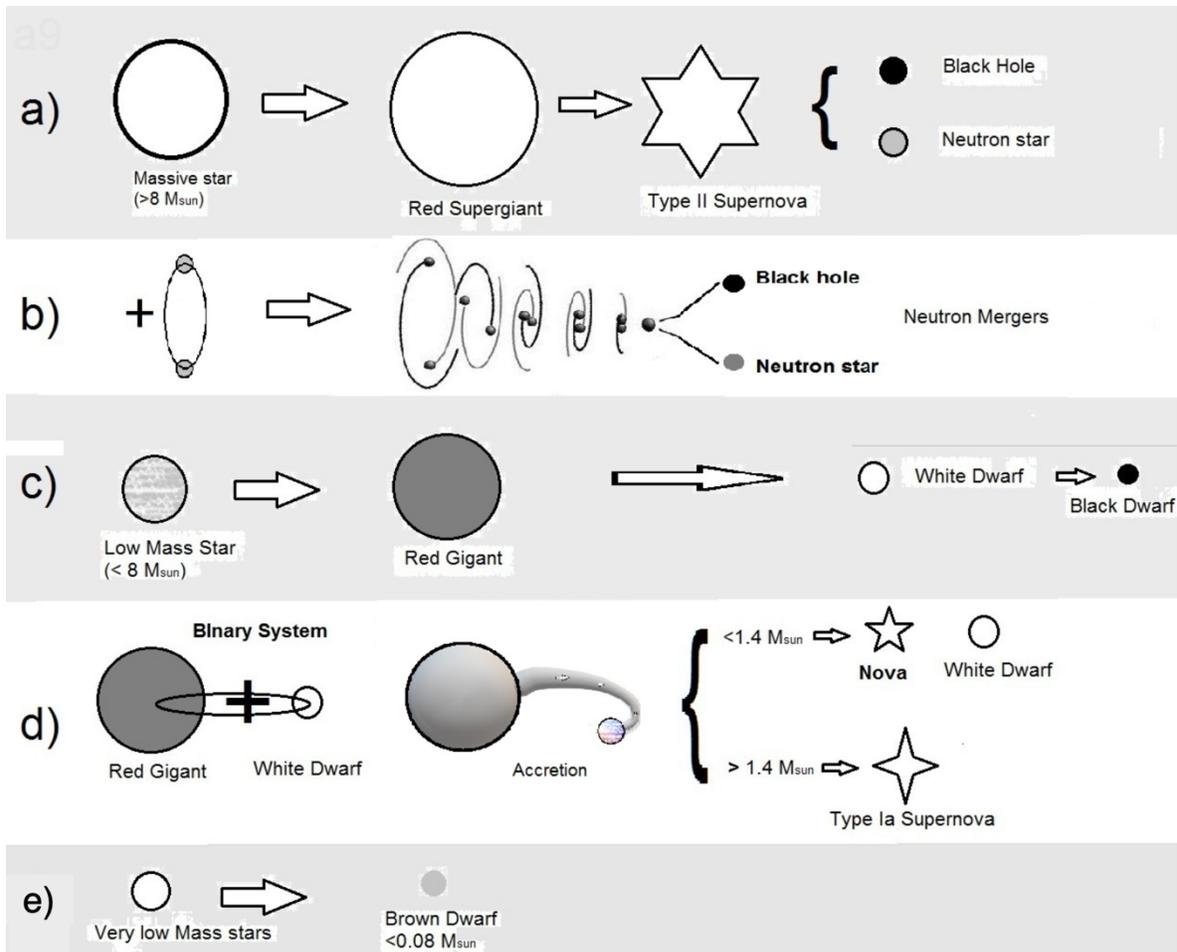

**Figure 9. Sketches of star evolution outcomes related to their mass:** a) Massive star. Heavy elements nucleosynthesis and ejection occurs during SNe explosions. b) Neutron stars bound together finished inspiraling and merged, creating the heaviest elements. c) Lower mass stars burn their fuel creating up to iron peak elements. d) Remnants of stars may pair together. Accretion occurs. Chandrasekar limit determines either a SnIa or a Nova explosion. Middle-table elements are created and spread into interstellar medium. e) very low mass stars end up as Brown dwarfs (figure credit, own source).

All the explosive processes that we have just described, in addition to the fission or rupture of already formed nuclei, rupture by very energetic cosmic rays (spallation), have been going on through eons of years. This has produced the abundance and distribution of chemical elements that we observe today. Finally we show the current table of elements in figure 10.

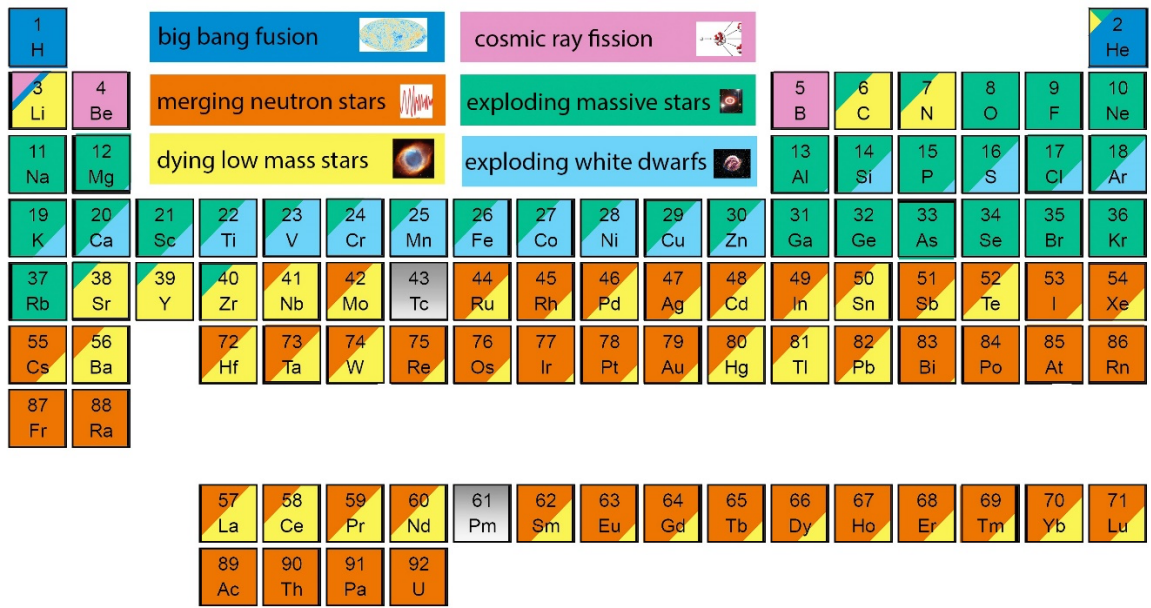

**Figure 10 Current 18-column form of the periodic table of elements 8 billion years after the Big Bang:** The process or processes that synthetized each one of the elements are indicated by the shadow intensity and mark in their respective boxes. Amounts are approximately proportional to shadows area (figure credit, after Johnson, 2020).

## 2. Expectations for the near future, a prospective:

We are at a unique period in the history of the human understanding where the vision on the origin of elements is beginning to clear out. There is a consensus that during the development of the universe there have been three stages in the nucleosynthesis of the elements. The first stage known as the standard Big Bang nucleosynthesis (SBBN) is related to the origin of the universe itself. The second stage begins with the recombination epoch and expands to the advent of the stelliferous era that overlaps in time with the end of the dark Ages, when sources of light appeared with the arrival of the first generation of stars (Pop III stars).

Pop III stars not only produced visible light, but also copious amounts of new elements, up to the iron peak. However, and although there is consensus agreement that all elements between C and the Fe peak have been produced by thermonuclear reactions inside stars, either, during their quiescent evolutionary stages or during the violent explosions (supernovae) that mark the deaths of some stars, there remain many questions concerning the typical mass scale of a Pop III star.

There has been a slow but steady paradigm shift concerning their typical mass scale of Pop III stars. From the acceptance of the past existence of very massive stars (above 100 $M_{sun}$) to computer simulations yielding stars with very much lower masses (of the order of few tens of a solar mass). This riddle might be solved in the near future with the James Webb Space Telescope (JWST) launched in 2021. If these Pop III stars were indeed such enormously huge objects, they might well be observed with the JWST in the not-too-distant future. Moreover, if it happens that Pop III stars are confirmed that they were gigantic, this will point them as the seeds of today galaxies' supermassive central black holes.

In a different vein, there are still some fine-tuning problems regarding r-process nucleosynthesis models. To test r-process models, nuclear physicists will need to obtain measurements or solid predictions of the fundamental properties of heavy, unstable nuclei that lie far from the valley of stability occupied by familiar long-lived isotopes. They'll need to know: masses, nuclear interaction cross sections, and decay rates. Procuring such data will be a primary science driver for research at several international accelerator facilities, among them the Facility for Rare Isotope Beams. This laboratory has been recently constructed at the campus of Michigan State University.

On the other hand, future nuclear physics predictions shall need to be consistent with imposed constraints -derived from astronomical observations- on the abundances of heavy elements. These predictions will be tested in the ultimate laboratory, the universe itself. In that regard, nature has shown scientists its kindness by wrapping the milky way with a halo of UFD satellite galaxies containing chemically primitive stars that exhibit high levels of r-process elements in particular, old and at the same time low metallicity stars. The study of this class of stars has just begun and it is expected to provide us with insightful answers on the origins of the abundance of its r-process elements.

On merging neutron stars of binaries, as we have already stated in this paper, the second event was detected in 2019 by the LIGO/Virgo collaboration. This detection is the second of hopefully many multi-messenger observations to come. These gravitational wave detections together with future high precision electromagnetic observations not only ground based but with space-borne instruments, might shed light on the long-standing questions about the neutron star internal structure and the supranuclear matter equation of state. In this respect, a current payload on board of the International Space Station (ISS) is devoted to the study of neutron stars through soft X-ray timing. This mission was recently launched on June 3, 2017. This mission was named the Neutron star Interior Composition Explorer (NICER) and it is expected to shed some light on the very nature of neutron stars.

On the other hand, the light and fragile isotopes of Li, Be, and B are not produced in stellar interiors (rather, they are destroyed at high temperatures), but by spallation reactions, with high-energy cosmic-ray particles removing nucleons from the abundant C, N, and O nuclei of the interstellar medium.

Lastly, the precise determination of the elements of the period table is very important for the whole history of the universe. The amount of the first elements or its isotopes (H, He, Li, D) determines, in the first minutes of the Universe infancy, details of the SBBN scenario: nuclei reactions rates, rate of expansion of the Universe, and the number of relativistic species present at that time. Details from SBBN can be finely constrained by measurements of the CMB acoustic peaks at the last scattering surface, that happened scarcely 380,000 years after the Big Bang. These acoustic peaks are measured today through anisotropies of the CMB at different angles in the sky. These depend very much on the amount of baryons (protons, neutrons) in the Universe, and finely on the amount of helium and lithium. Data from Planck satellite helped to determine the amount of baryonic matter to a precision of 0.9% (Fields, 2020). Related to this is also the number of leptons families (relativistic neutrinos) allowed by CMB data; this in turn is sensible to the helium and lithium abundances.

However, the "Schramm plot" shown in figure 3, clearly shows that Li/H measurements are inconsistent with the D/H obtained amounts (and CMB), given the error bars indicated there. In addition, recent updates in nuclear cross sections and new revisions of stellar abundance systematic errors, increase the discrepancy to over 5σ, depending on the stellar abundance analysis adopted (Iocco, 2009). And what is more, latest predicted primordial abundances - the most accurate to date- for D, $^3$He and $^4$He are in agreement with observations, within error bars. But for $^7$Li, the situation is quite different as abundance calculations remain a factor of ≈3 above the primordial abundance level of lithium (the Spite plateau (Spite, 1982)). See Pitrou paper for a comprehensive discussion (Pitrou, 2018). This great discrepancy has been called "the lithium problem".

In future, in about a decade, finer CMB probes, the so called CMB-Stage 4, will reveal these details to confirm or dismiss current scientific knowledge. One of the key issues is the lithium problem mentioned, that still remains, posing an important challenge to theoretical nuclei reaction rates and data consistency. Also, in near future joint data analysis of CMB probes with those of new galaxy surveys, such as DESI (Dark Energy Spectroscopic Instrument), will be able to determine the number of relativistic degrees of freedom (neutrino families) and sum of the mass of neutrinos, will help to constrain the helium abundance in the Universe and number of neutrinos families.

Currently it is considered that the "lithium problem" mismatch could be due to either: systematic errors in the observed abundances, and/or uncertainties in stellar astrophysics including stellar depletion or nuclear inputs, or possibly some new physics beyond the Standard Model (Fields, 2011). Although the Standard Model provides a precise description of physics up to the Fermi scale, cosmology cannot be traced in detail before the Big Bang.

Today, Big Bang Nucleosynthesis indicates where lays the boundary, between the established and the speculative cosmology. At present is uncertain how it would be possible to push this frontier back to the quark-hadron transition, or electroweak symmetry breaking as so far no relics of these epochs have been observed.

In the second decade of the 21st century we have advanced a great deal in understanding the cosmological origin of the elements. It is said that upcoming years belong to an era of joint probes analysis, measuring light from CMB, from galaxy clustering, from individual astrophysical objects and, last not least, gravity waves originated from neutron stars collapse, to combine information and to gain in constraining theoretical parameters. Soon there will be new space and ground-based observatories and a next generation of nuclear accelerators. May be present questions will be answered but new uncertainties will arise.

Future is near, but we have to put hands on now!

## Acknowledgement

JLCC acknowledges support from CONACyT project 283151.